\documentclass[twocolumn,
showpacs,
preprintnumbers,
amsmath,
amssymb,
amsfonts,
final,
a4paper,
aps,
citeautoscript,
footinbib,
prl,
superscriptaddress]{revtex4}
\usepackage{times}
\usepackage{graphicx}
\usepackage[latin1]{inputenc}

\begin{document}
\title{Quantum criticality in
the SO(5) bilinear-biquadratic Heisenberg 
chain}
\author{F. Alet}
\affiliation{Laboratoire de
Physique Th\'eorique, CNRS UMR 5152,
Universit\'e Paul Sabatier, F-31062 Toulouse, France}
\author{S.\ Capponi} \affiliation{Laboratoire de
Physique Th\'eorique, CNRS UMR 5152,
Universit\'e Paul Sabatier, F-31062 Toulouse, France}
\author{H. Nonne}
\affiliation{Laboratoire de Physique Th\'eorique et
Mod\'elisation, CNRS UMR 8089,
Universit\'e de Cergy-Pontoise, Site de Saint-Martin,
F-95300 Cergy-Pontoise Cedex, France}
\author{P. Lecheminant}
\affiliation{Laboratoire de Physique Th\'eorique et
Mod\'elisation, CNRS UMR 8089,
Universit\'e de Cergy-Pontoise, Site de Saint-Martin,
F-95300 Cergy-Pontoise Cedex, France}
\author{I. P. McCulloch}
\affiliation{School of Physical Sciences, The University of Queensland, Brisbane, QLD 4072, Australia}

\date{\today}

\begin{abstract}
The zero-temperature properties of the SO(5)
bilinear-biquadratic Heisenberg spin chain
are investigated by means of a low-energy
approach and large scale numerical calculations.
In sharp contrast to the spin-1 SO(3) Heisenberg chain,
we show that the SO(5) Heisenberg spin chain is dimerized
with a two-fold degenerate ground state.  On top of this
gapful phase, we find the emergence of a non-degenerate
gapped phase with hidden (Z$_2$ $\times$ Z$_2$)$^2$ symmetry
and spin-3/2 edge states that can be understood from a
SO(5) AKLT wave function. We derive a low-energy theory
describing the quantum critical point which separates
these two gapped phases. It is shown and confirmed numerically
that this quantum critical point belongs to
the SO(5)$_1$ universality class.
\end{abstract}

\maketitle
One-dimensional (1D) quantum systems display a wealth of
fascinating behaviors which have attracted much interest
over the years.
A paradigmatic example of an exotic phase stabilized
by 1D quantum fluctuations
is the Haldane phase of the spin-1 SU(2) Heisenberg chain
\cite{haldane}.
On top of the existence of a gap, this phase displays remarkable properties like
the existence of a hidden N\'eel antiferromagnetic order~\cite{dennijs} or
the emergence of fractional spin-1/2 edge states
when the chain is doped by non-magnetic impurities~\cite{hagiwara}.
A simple way to grasp the main characteristics
of the Haldane phase is provided by the seminal work of Affleck, Kennedy, Lieb andTasaki (AKLT)
~\cite{aklt}, where a fine-tuned biquadratic
exchange interaction is introduced so that the ground state (GS) is made up solely of nearest-neighbor valence
bonds. Due to the absence of
a quantum phase transition upon switching on the biquadratic exchange
from the Heisenberg model up to the AKLT point, all the exotic properties of the Haldane phase can be simply deduced from the AKLT wave function~\cite{jolicoeur}.

In an interesting work, Tu {\it et al.}~\cite{Tu2008} have
recently studied a generalization
of the spin-1 bilinear-biquadratic Heisenberg chain to
higher symmetry group SO(2$n$+1), with Hamiltonian:
\begin{equation}
 {\cal H} = \cos\theta \sum_{i} \sum_{a<b} L^{ab}_i L^{ab}_{i+1}
+ \sin \theta \sum_{i} (\sum_{a<b} L^{ab}_i L^{ab}_{i+1})^2,
\label{model}
\end{equation}
where $L_{ab}$ ($1\leq a<b \leq 2n+1$) are the $n(2n+1)$ generators which
transform in the vectorial representation ($n \times n$ matrices) of the
SO(2$n$+1) group. They are normalized such that the single-site Casimir
operator is: $\sum_{a<b} (L^{ab}_i)^2 = 2n$.
For $n=1$, this model is nothing but the spin-1 bilinear biquadratic Heisenberg chain.
The phase diagram of the model (\ref{model}) for general $n$ has been conjectured
in Ref.~\onlinecite{Tu2008} based on the behaviour at some special points.
For $\theta=\tan^{-1} 1/(2n-1)$, the model is the SU(2$n$+1) Sutherland model
which displays a quantum critical behavior with
$2n$ gapless modes \cite{sutherland-affleck88}.
There is a second SU(2$n$+1) symmetric model for $\theta = \pm \pi/2$ with
alternating fundamental and conjugate representations, where the GS is
dimerized with a spontaneous breaking
of translation symmetry \cite{affleck91}.
For $\theta_{AKLT}= \tan^{-1} 1/(2n+1)$, the GS is a SO(2$n$+1)
matrix product state which is the generalization of the SO(5) AKLT
state introduced by Scalapino {\it et al.} \cite{szh} in the context of a SO(5) two-leg ladder.
The resulting non-degenerate gapped phase is the SO(2$n$+1) generalization of the
Haldane phase with hidden antiferromagnetic
ordering 
due to the breaking of a $(Z_2 \times Z_2)^n$ symmetry
\cite{Tu2008}.
In open boundary geometry, the two edge states belong to the spinorial
representation of SO(2$n$+1) with dimension $2^n$, so that the GS is $4^n$-fold degenerate. Finally, the model with
 $\theta_R = \tan^{-1} (2n-3)/(2n-1)^2$ is
exactly solvable~\cite{Reshetikhin1985} and is expected
to have gapless excitations from general grounds.

In this paper, we investigate part of the zero-temperature phase
diagram between $\theta=0$ and $\theta_{AKLT}$, including the criticality
of the SO(5) bilinear-biquadratic spin chain by means of complementary
low-energy approach and several numerical calculations: exact diagonalization (ED)
and density matrix renormalization group (DMRG) \cite{DMRG}.
We give further arguments supporting the phase diagram for $n=2$ conjectured in Ref. \onlinecite{Tu2008}.
In sharp contrast with the spin-1 Heisenberg chain, the AKLT point for $n=2$ is found {\it not} to capture the physics
of the SO(5) Heisenberg chain. A quantum phase transition
occurs at $\theta = \theta_R$ that we fully characterize 
and show to belong to the
SO(5)$_1$ universality class with central charge $c=5/2$.

{\it Low-energy approach.}
In principle, a field-theory approach capturing the low-energy properties of the
SO(2$n$+1) Heisenberg spin chain can be derived 
from
the solvable point at $\theta = \theta_R$. By exploiting the integrability of the model, one may determine
the nature of the underlying conformal field theory (CFT) and investigate
small deviations $|\theta - \theta_R| \ll 1$ in parallel to the Majorana fermion
approach of Tsvelik for $n=1$ \cite{tsvelik90}.
Here, we instead directly derive a low-energy description for the SO(5) Heisenberg spin chain
by considering a spin-3/2 fermionic model which is SO(5) symmetric
\cite{zhang2003,Wurev}:
\begin{eqnarray}
{\cal H}_{SO(5)} &=& -t \sum_{i,\alpha} [c^{\dagger}_{\alpha,i} c_{\alpha,i+1} +
{\rm H.c.} ]
- \mu \sum_i n_i
\nonumber \\
&+& \frac{U}{2} \sum_i n_i^2 + V \sum_i P^{\dagger}_{00,i}
P_{00,i},
\label{so5model}
\end{eqnarray}
where $c^{\dagger}_{\alpha,i}$ denotes the fermionic creation operator
with spin index $\alpha = \pm 3/2, \pm  1/2$, and
$n_i = \sum_{\alpha}
c^{\dagger}_{\alpha,i} c_{\alpha,i}$ is the density operator
at site $i$.
In Eq. (\ref{so5model}),
the singlet BCS pairing operator for spin-3/2 fermions is
$P^{\dagger}_{00,i} =
c^{\dagger}_{3/2,i} c^{\dagger}_{-3/2,i} - c^{\dagger}_{1/2,i} c^{\dagger}_{-1/2,i}$.
As shown in Ref.\onlinecite{zhang2003}, the spin-3/2 model (\ref{so5model})
enjoys a U(1)$_{\rm charge}$ $\times$ SO(5)$_{\rm spin}$ continuous symmetry without
any fine-tuning.
When $U,V >0$, the lowest states for $t=0$ at half-filling are the quintet states:
\begin{center}
\includegraphics[width=0.99\columnwidth]{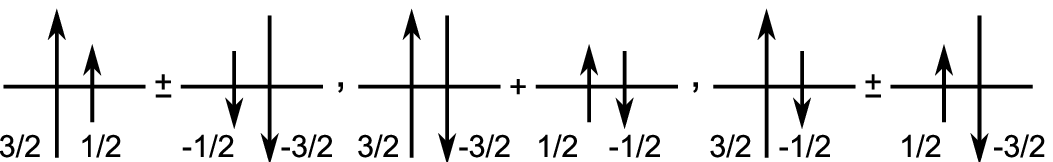}
\end{center}
A standard strong-coupling approach can then be applied when $U,V \gg t >0$ and
one finds the effective model to be a SO(5) Heisenberg spin  chain:
${\cal H} = J \sum_{i} \sum_{a<b} L^{ab}_i L^{ab}_{i+1}$ with $J = 2 t^2/3U$.
Then, a low-energy description can be deduced for the latter model given the adiabatic continuity between weak and strong coupling regimes for the spin-3/2 model~\cite{heloise}.
As derived in Ref.~\onlinecite{heloise}, the weak-coupling approach at half-filling  is built from eight right and left moving Majorana fermions $\xi_{R,L}^{A}$ (with $A=1,\ldots,8$):
\begin{eqnarray}
{\cal H}_{\rm int} &=&
\frac{g_1}{2} \; \left(\sum_{a=1}^{5} \xi_R^{a} \xi_L^{a} \right)^2
+ g_2 \; \xi_R^{6} \xi_L^{6}
\sum_{a=1}^{5} \xi_R^{a} \xi_L^{a}
\nonumber \\
&+& \frac{g_3}{2} \; \left(\xi_R^{7} \xi_L^{7}
+ \xi_R^{8} \xi_L^{8}  \right)^2
\nonumber \\
&+& \left( \xi_R^{7} \xi_L^{7} + \xi_R^{8} \xi_L^{8} \right)
\left( g_4 \;
\sum_{a=1}^{5} \xi_R^{a} \xi_L^{a}
+
g_5 \; \xi_R^{6} \xi_L^{6} \right),
\label{majofthalfilling}
\end{eqnarray}
where the two Majorana fermions $\xi^{7,8}_{R,L}$ account for the
U(1) charge symmetry, the five Majorana fermions $\xi^{a}_{R,L}$ (with $a = 1, \ldots, 5$)
generate the SO(5) symmetry, and $\xi^{6}_{R,L}$
describes an internal discrete Z$_2$ symmetry.
By integrating out the charge degrees of freedom
which are fully gapped in the large U,V repulsive limit, one then obtains an effective Hamiltonian with leading part
\begin{eqnarray}
{\cal H}_{\rm eff} = - \frac{i v}{2} \sum_{a=1}^{5}
\left( \xi^a_R \partial_x \xi^a_R - \xi^a_L \partial_x \xi^a_L \right)
- i m_s \sum_{a=1}^{5} \xi^a_R \xi^a_L,
\label{effectiveHam}
\end{eqnarray}
describing five massive Majorana fermions with mass $m_s \sim U > 0$.
In fact, after the integration of the charge
degrees of freedom, $\xi^{6}_{R,L}$ is also a massive Majorana fermion,
but with a higher mass $m_o > m_s$ so that its contribution can be
neglected in the low-energy limit $E \ll m_0$.
The effective low-energy approach of the
SO(5) Heisenberg spin chain is thus given by model (\ref{effectiveHam}),
a natural extension to SO(5) of Tsvelik's result (i.e. three massive Majorana
fermions) for the spin-1 Heisenberg chain \cite{tsvelik90}.
Since a 1D theory of massive Majorana accounts for the long-distance behavior
of a 1D quantum Ising model, the low-energy properties of the
SO(5) Heisenberg spin chain can be described in terms of
five decoupled off-critical 1D quantum Ising models
in their disordered phases ($m_s \sim T -T_c$).

To fully characterize the nature of the GS of the SO(5) Heisenberg
chain, it is instructive to investigate the expectation
value of the SO(5) dimerization operator $D_i = (-)^i  \sum_{a<b} L^{ab}_i L^{ab}_{i+1} $.
One can follow a similar strategy to derive a low-energy expression
for this operator. In the low-energy limit $E \ll m_o$, we find that $D$ is expressed
in terms of the disorder operators $\mu_a$ of the underlying
1D quantum Ising degrees of freedom:
$D \sim \prod_{a=1}^{5} \mu_a$. Since the Ising models are 
locked in their disordered phases, we obtain that $\langle D \rangle \ne 0$.
The GS of the SO(5) Heisenberg spin chain is thus two-fold degenerate
and spontaneously dimerized in sharp
contrast with the Haldane phase of the spin-1 Heisenberg chain.

With this low-energy approach at hand, we can investigate the effect
of the SO(5) biquadratic term in Eq. (\ref{model}) in the vicinity of $\theta =0$. The most relevant term consistent with
the symmetries of the lattice model is the Majorana fermion mass term of Eq.~(\ref{effectiveHam}).
We thus deduce that the low-energy Hamiltonian of Eq.~(\ref{model})
is still given by (\ref{effectiveHam}) with a mass $m_s(\theta)$,
a phenomenological parameter within our approach.
The only possibility is that the mass $m_s$ vanishes and changes
sign to form a new gapful phase with the five effective
1D quantum Ising models entering their ordered phases.
It is very tempting to identify the quantum critical point at $m_s =0$
with the location of the integrable model at $\theta=\theta_R$
which is expected, on general grounds, to display a quantum critical
behavior. This will be confirmed precisely numerically below.
We thus conclude that the universality class of the integrable
model for $\theta = \theta_R$ is the SO(5)$_1$ CFT with central
charge $c=5/2$.
In addition, our low-energy approach enables us to extract
the leading asymptotics of the SO(5) spin-spin correlations
which display a power-law decay with a universal exponent:
$\langle L^{ab}_{i+x} L^{ab}_{i} \rangle
\sim (-)^{x} x^{-5/4}$. The SO(5) dimerization operator
vanishes at $\theta = \theta_R$ and has a power-law decaying
correlation function with the same universal exponent.
For $\theta > \theta_R$, a new gapful phase emerges with
no spontaneous dimerization since $\langle \mu^a \rangle = 0$
for $m_s <0$. This phase is the analog of the Haldane phase of the spin-1 Heisenberg
chain. Each Ising model has a doubly degenerate GS which
gives a $2^{5}$ degeneracy. However, there is a redundancy in the Majorana
fermion description since the transformation $\xi^a_{R,L} \rightarrow
- \xi^a_{R,L}, \mu^a \rightarrow \mu^a, \sigma^a \rightarrow - \sigma^a$
leaves invariant the Hamiltonian and the physical operators.
The correct hidden symmetry group is thus $(Z_2 \times Z_2)^2$ in full
agreement with the description of the AKLT wave function for the
SO(5) generalization of the Haldane phase \cite{Tu2008,szh}.
As a consequence of this antiferromagnetic ordering,
there are non-trivial edge states in this phase when
a semi-infinite open chain is considered.
Using the results of Ref. \onlinecite{orignac}, we find the emergence
of five localized Majorana fermion zero mode states $\eta^a$ inside
the gap (mid-gap states).
These five local fermionic states give rise to a local spin-3/2 operator
${\cal \vec S}$ since:
\begin{eqnarray*}
{\cal S}^x &=& -i \eta^1 \eta^4 - i \eta^2 \eta^5 - i \sqrt{3} \; \eta^2 \eta^3
\nonumber \\
{\cal S}^y &=& -i \eta^1 \eta^5 + i \eta^2 \eta^4 + i   \sqrt{3}  \; \eta^1 \eta^3
\nonumber \\
{\cal S}^z &=& -i \eta^1 \eta^2 - 2 i \eta^4 \eta^5 ,
\end{eqnarray*}
describes a local spin-3/2 operator thanks to the anticommutation
relations of the Majorana fermions: $\{\eta^a, \eta^b \} = \delta^{ab}$.
In the phase with $\theta > \theta_R$, we thus predict the occurence
of spin-3/2 boundary excitations at the edge of the chain, again in full agreement with the SO(5) AKLT wave-function \cite{szh}.

{\it Numerical results.}
We combine ED and DMRG calculations to clarify the phase diagram and the criticality of the phase transition. We use a spin-2 formulation of the model (\ref{model}) using projectors~\cite{Tu2008}, which leads to
\begin{widetext}
\begin{equation}
{\cal H} = \sum_{\langle ij\rangle} \cos\theta \bigl[ -1 - \frac{5}{6} {\bf S}_i \cdot {\bf S}_j
+ ({\bf S}_i \cdot {\bf S}_j)^2/9 +  ({\bf S}_i \cdot {\bf S}_j)^3/18 \bigr] + \sin\theta \bigl[ 1 - 5 {\bf S}_i \cdot {\bf S}_j - \frac{17}{12} ({\bf S}_i \cdot {\bf S}_j)^2 + \frac{1}{3}({\bf S}_i \cdot {\bf S}_j)^3 + \frac{1}{12}({\bf S}_i \cdot {\bf S}_j)^4 \bigr].
\label{spin2model.eq}
\end{equation}
\end{widetext}

\begin{figure}[!ht]
\includegraphics[width=\linewidth,clip]{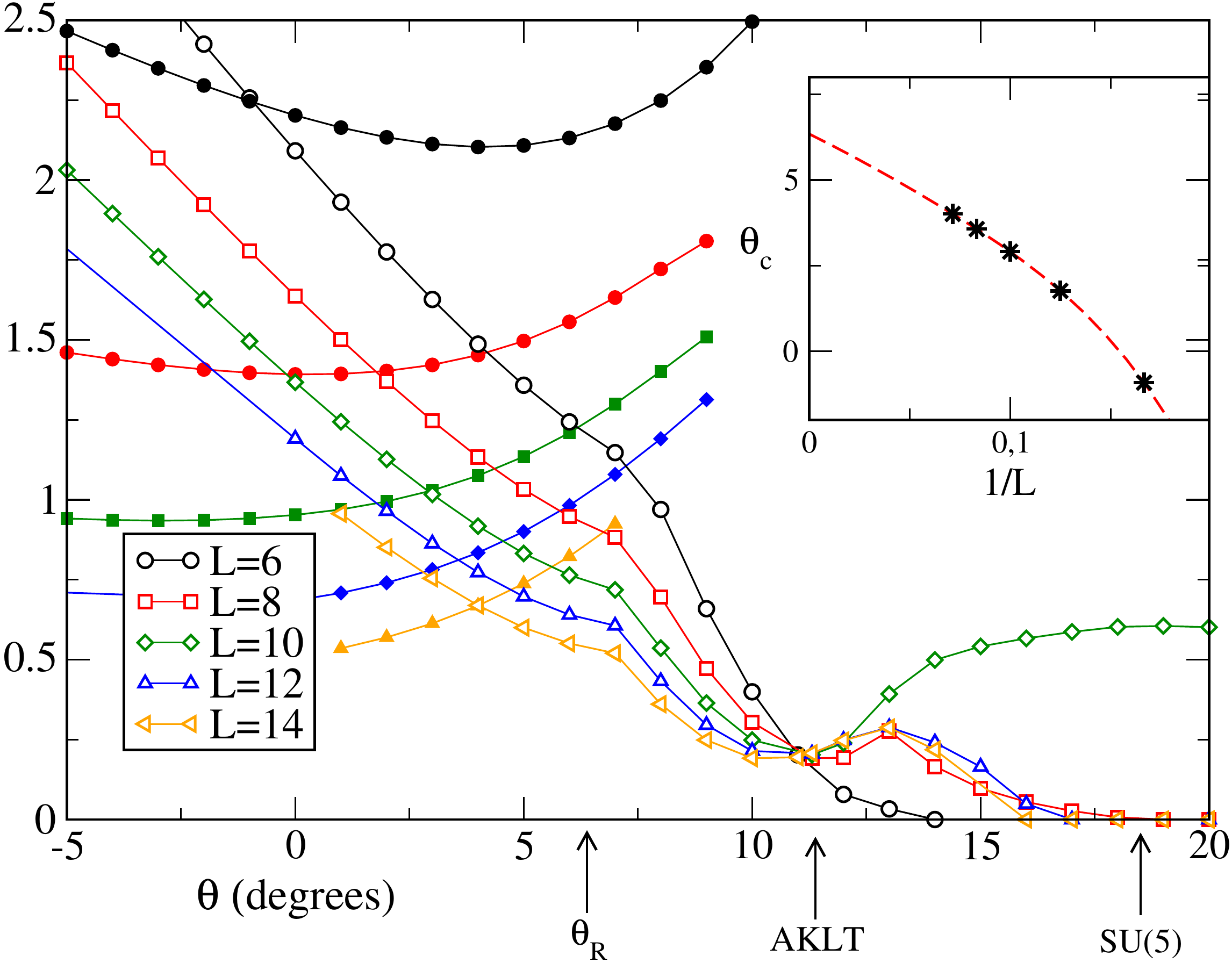}
\caption{ED data showing the gap to the first singlet and magnetic excitations (empty and filled symbols respectively) vs $\theta$ for several chain lengths $L$. Inset shows the extrapolation of the crossing point.}
\label{fig:phase_diag_ed}
\end{figure}

First, we compute with ED the first singlet and magnetic excitations for various chains with periodic boundary conditions (PBC). For $\theta< \theta_R$, the data (shown in Fig.~\ref{fig:phase_diag_ed}) are compatible with a finite spin gap but a vanishing singlet gap (with momentum $\pi$), which suggests a singlet phase that breaks translation symmetry. It is remarkable that the extrapolation (using polynomial fit) of the crossing point of the first magnetic and non-magnetic excitations gives $\theta_c=6.34^\circ$ which is precisely the numerical value of $\theta_R$.

\begin{figure}[!ht]
\includegraphics[width=\linewidth,clip]{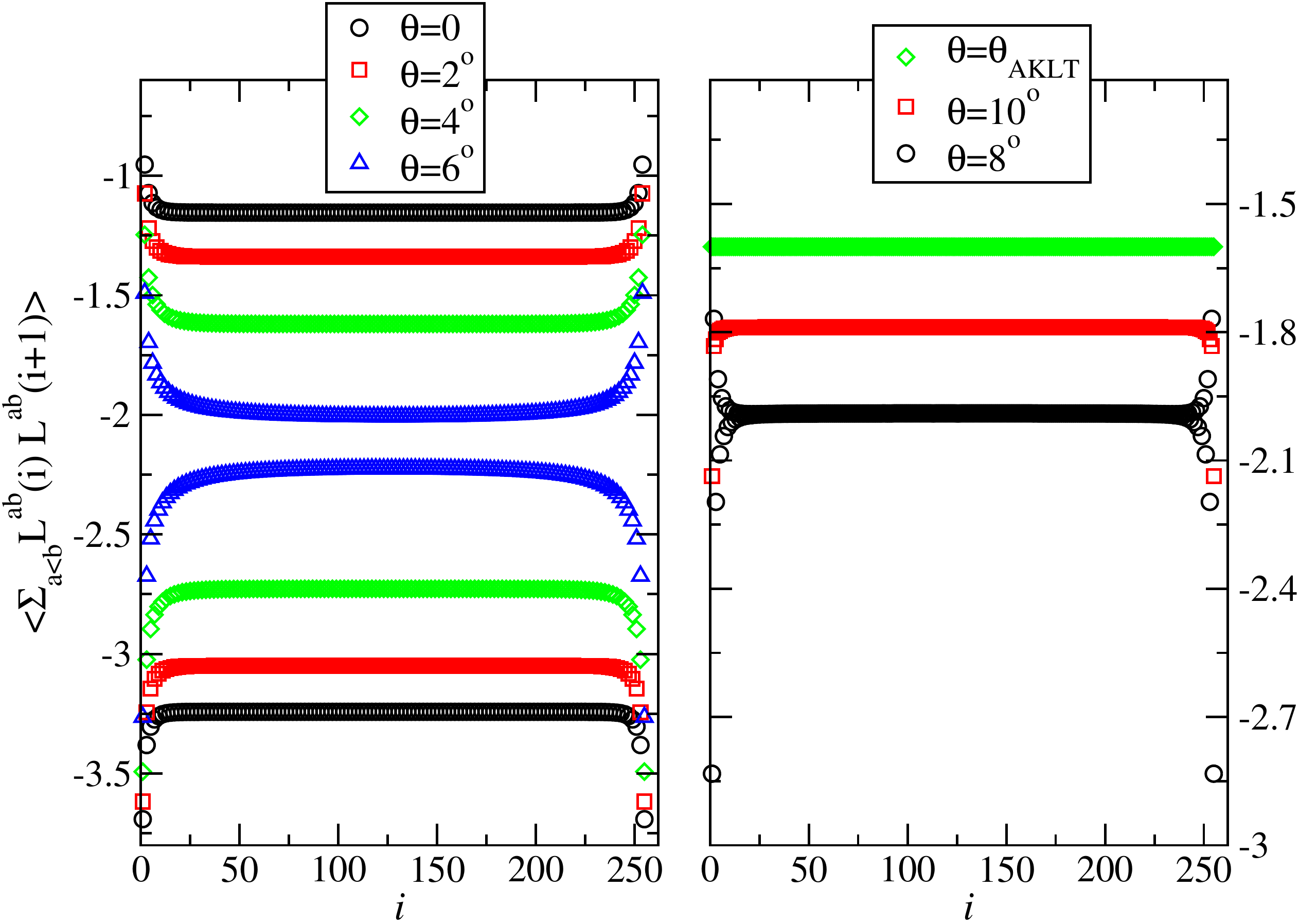}
\caption{DMRG data of the dimerization $\langle  \sum_{a<b} L^{ab}_i L^{ab}_{i+1} \rangle$ as a function of position $i$, induced in a $L=256$ chain by the OBC in the ranges $\theta < \theta_R$ (left panel) and $\theta>\theta_R$ (right panel). At $\theta=0$, the system is strongly dimerized.}
\label{fig:dim}
\end{figure}

The dimerized nature of the GS of the Heisenberg SO(5) chain at $\theta=0$ is confirmed by computing the correlator $\langle  \sum_{a<b} L^{ab}_i L^{ab}_{i+1} \rangle$ as a function of position $i$ in an open chain. The DMRG data (see Fig.~\ref{fig:dim}) obtained on a $L=256$ chain with open boundary conditions (OBC) in the range $\theta \in [ 0,14^\circ]$ clearly show a staggered pattern for $0 \leq \theta < \theta_R$, in contrast with the uniform values found for $\theta > \theta_R$, for instance at the AKLT point.

For $\theta \simeq \theta_{AKLT}$, the lowest excitation is magnetic and has a finite gap. In fact, right at AKLT point, ED
\emph{proves} the existence of a finite gap in the thermodynamic limit using an argument from Knabe~\cite{Knabe1988}:
the thermodynamic gap is bounded by the gap on a finite cluster with $L+1$ sites
(OBC) as $ \Delta_\infty > \frac{L}{L-1} (\Delta_L-\frac{1}{L})$.
 Using $L=8$, we already obtain that the spin gap is finite and satisfies $\Delta_\infty > 0.185$.

 For larger $\theta$, data are also compatible with a SU(5) gapless phase but we
 observe strong finite-size effects due to incommensurate excitations, similar to the case of the spin-1 chain close to SU(3) point~\cite{Golinelli1999}.

\begin{figure}[!ht]
\includegraphics[width=\linewidth,clip]{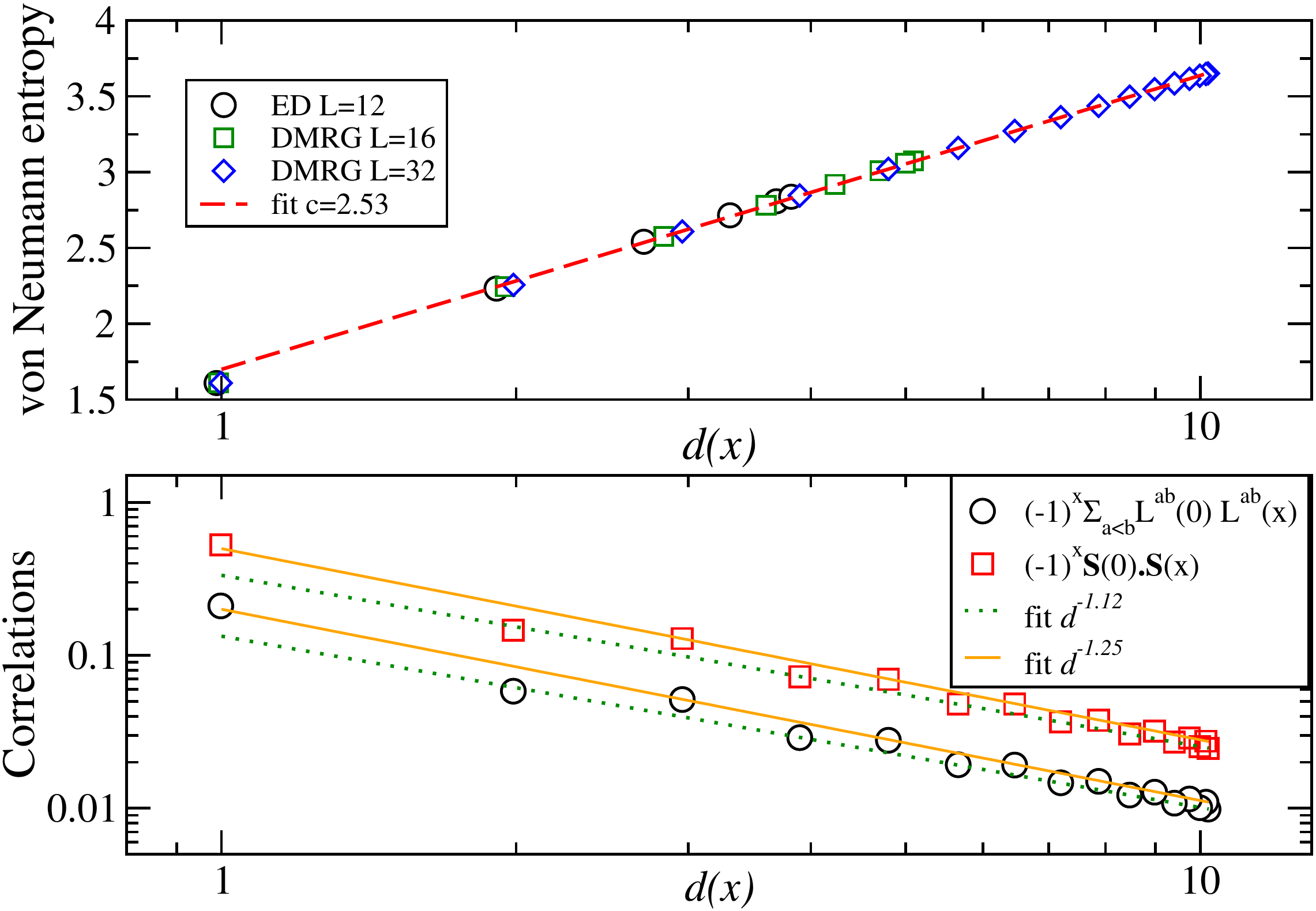}
\caption{At $\theta_R$ and as a function of  conformal distance $d(x)=L/\pi \sin (x\pi/L)$ for systems with PBC: (a) von Neumann entropy of a block of size $x$ (log-lin scale), (b) Spin $ \langle(-)^x{\bf S}_0\cdot {\bf S}_x \rangle $ and SO(5) generators $\langle (-)^x  \sum_{a<b} L^{ab}_0 L^{ab}_x \rangle $ correlators (log-log scale). Correlators are translation-invariant due to the PBC and are normalized to $1$ at their onsite values.}
\label{fig:entropy-correl}
\end{figure}

As argued in the low-energy section, the central
charge of the critical point between the dimerized and SO(5) Haldane phases should be $c=5/2$, corresponding to a
SO(5)$_1$ CFT. In Fig.~\ref{fig:entropy-correl}a, we show in log-linear scale the von Neumann entropy $S(x)$ of a block of $x$ sites at $\theta_R$ as a function of the conformal distance $d(x)=L/\pi \sin (\pi x/L)$, for chains with PBC of sizes up to $L=32$. Fitting the data to the expected form for PBC $S(x)=c/3
\ln(d(x))+K$ with $K$ a constant (see Ref.~\onlinecite{Cardy}), we obtain an estimate of the central charge $c=2.53$. We can also obtain $c$ using finite-size corrections of the GS energy per site: $e_0(L) = e_0 - (\pi v c)/6L^2$. Using ED data (not shown) for both the energy and finite-size velocity $v(L)=E_0(k=2\pi/L)-E_0(k=0)$,  we obtain the same estimate $c=2.53$, in excellent agreement with a SO(5)$_1$ quantum criticality.

At the critical point $\theta_R$, the low-energy approach predicts the leading behavior of two-point correlations:
$\langle L^{ab}_i L^{ab}_{i+x} \rangle \sim (-)^{x} x^{-5/4}$ (the same dependence is expected for spin correlations $\langle {\bf S}_i\cdot {\bf S}_{i+x}\rangle$). These correlators, plotted in Fig.~\ref{fig:entropy-correl}b for a $L=32$ chain with PBC, exhibit an even/odd effect besides the $(-)^x$ factor. Fits of the two sets of points respectively lead to power-law exponents of $1.25$  and $1.12$, equal or close to the theoretical prediction $5/4$ for both correlators. We expect that all correlators decay with the same exponent $5/4$ in the thermodynamic limit and that the even/odd effect (and the corresponding different exponents) is caused by subleading correlations, which can be important for the moderate system size $L=32$ that can be reached in the DMRG simulations.

{\it Conclusion.}
Using complementary techniques, we
confirm the phase diagram that has been proposed phenomenologically
in Ref.~\onlinecite{Tu2008}.
In particular, we derive a low-energy approach in terms of five massive
Majorana fermions
which allows to determine that (i) the GS at $\theta=0$
is spontaneously dimerized and (ii) it is separated from the
SO(5) generalization of the Haldane phase
by a quantum critical point described by a SO(5)$_1$ CFT
with  central charge $c=5/2$.
All these predictions are quantitatively confirmed
using extensive numerical simulations.
In sharp contrast to the $n=1$ case,
the SO(5) AKLT point does not share the same physics with the
Heisenberg point $\theta =0$.
This property should be a general feature of the model with $n >1$ \cite{Tu2008}.
In this respect, the $n=1$ case is very special
and {\it not} representative of the generic situation.

To give some perspective, let us mention that the low-energy approach
presented here is directly relevant to the quantum critical behavior
of the spin-2 model interpolating between two topologically
distinct AKLT states \cite{Zang2010,dhlee,Jiang2010}.
In particular, the quantum phase transition between the SO(5) AKLT
state and the spin-2 dimerized phase found in Refs. \onlinecite{Zang2010,Jiang2010}
can be shown to be described by the effective Hamiltonian (\ref{effectiveHam})
and thus belongs to the SO(5)$_1$ universality class.
However, our approach does not capture the direct topological quantum phase transition
between the spin-2 AKLT state with spin-1 edge states and the SO(5) AKLT state
put forward in Refs. \onlinecite{Zang2010,dhlee,Jiang2010}.
It will be interesting to understand how it can
be extended to explain the emergence of SO(5)$_1$ criticality recently observed numerically for this transition \cite{Jiang2010}.

We thank E. Boulat  for useful discussions. Numerical simulations were performed using HPC resources from GENCI-CCRT (Grant 2010-x2010050225) and CALMIP.


\begin{thebibliography}{99}
\bibitem{footnote:DMRG} We use up to $m=1200$ SU(2) states, corresponding to about $10000$ usual U(1) states, for systems with PBC, with a long warm-up procedure performing at least 10 sweeps each time $m$ is increased by 50. Truncation error per site is at most $3.10^{-8}$.
\end{thebibliography}
\end{document}